\documentclass[12pt]{article}
\usepackage[margin=1in]{geometry}
\usepackage{graphicx}
\graphicspath{{figures/}}
\usepackage{amsmath,amssymb}
\usepackage{natbib} 
\usepackage{hyperref}
\usepackage{overpic}
\usepackage{caption}
\usepackage{subcaption}
\usepackage{subcaption}
\usepackage{amsmath}
\usepackage{url}
\usepackage{xcolor}

\makeatletter
\renewcommand\section{\@startsection
  {section}{1}{0mm}
  {\baselineskip} 
  {0.5em}         
  {\normalfont\normalsize\bfseries}}
\renewcommand\subsection{\@startsection
  {subsection}{2}{0mm}
  {\baselineskip}
  {0.5em}
  {\normalfont\normalsize\bfseries}}
\renewcommand\subsubsection{\@startsection
  {subsubsection}{3}{0mm}
  {\baselineskip}
  {0.5em}
  {\normalfont\normalsize\bfseries}}
\makeatother

\begin{document}

\title{High-resolution probabilistic estimation of three-dimensional regional ocean dynamics from sparse surface observations}

\author{
Niloofar Asefi$^{1}$ \and
Tianning Wu$^{3}$ \and
Ruoying He$^{3}$ \and
Ashesh Chattopadhyay$^{2,*}$
}

\date{}

\maketitle

\begin{center}
$^1$Department of Electrical and Computer Engineering, University of California, Santa Cruz, CA 95064, USA\\
$^2$Department of Applied Mathematics, University of California, Santa Cruz, CA 95064, USA\\
$^3$Department of Marine, Earth, and Atmospheric Sciences, North Carolina State University, Raleigh, NC 27695, USA\\[6pt]
\texttt{*aschatto@ucsc.edu}
\end{center}

\begin{abstract}
The ocean interior regulates Earth’s climate but remains sparsely observed due to limited in situ measurements, while satellite observations are restricted to the surface. We present a depth-aware generative framework for reconstructing high-resolution three-dimensional ocean states from extremely sparse surface data. Our approach employs a conditional denoising diffusion probabilistic model (DDPM) trained on sea surface height and temperature observations with up to 99.9\% sparsity, without reliance on a background dynamical model. By incorporating continuous depth embeddings, the model learns a unified vertical representation of the ocean states and generalizes to previously unseen depths. Applied to the Gulf of Mexico, the framework accurately reconstructs subsurface temperature, salinity, and velocity fields across multiple depths. Evaluations using statistical metrics, spectral analysis, and heat transport diagnostics demonstrate recovery of both large-scale circulation and multiscale variability. These results establish generative diffusion models as a scalable approach for probabilistic ocean reconstruction in data-limited regimes, with implications for climate monitoring and forecasting.
\end{abstract}

\vspace{2pc}

\section{Introduction}
Understanding ocean interior properties is crucial for understanding climate dynamics, as more than 90\% of Earth’s excess heat is absorbed by the oceans, where it is redistributed through ocean circulation and affects climatic variability \cite{trenberth2013hiatus, trenberth2010tracking, trenberth2015hiatus, yan2016hiatus, kosaka2013hiatus}.
Subsurface ocean observations are primarily obtained from in situ platforms, such as moorings, research vessels, and autonomous profiling systems. While these instruments provide high-fidelity measurements of vertical structure, their sampling is inherently sparse, irregular, and non-synoptic in space and time. Observations are collected along drifting trajectories or at isolated locations, with limited coverage across depth and basin-scale domains, preventing the direct construction of spatially resolved three-dimensional fields. As a result, the ocean interior is fundamentally under-observed, and its full state must be inferred from incomplete and heterogeneous measurements~\cite{dickey2003emerging,ingleby2007quality,riser2016fifteen,roemmich2000argo,gaillard2009quality}. The challenge is not just the data quality, but the geometry of the data; i.e., observations sample the ocean along sparse, irregular trajectories rather than as a continuous three-dimensional field. 

Satellite remote sensing has transformed physical oceanography by providing spatially extensive observations of surface variables, including sea surface height (SSH), sea surface temperature (SST), and sea surface salinity (SSS), enabling the characterization of mesoscale and submesoscale dynamics and their role in ocean circulation \cite{durand2010swot, fu2014transition, wang2019longwavelength, su2018submesoscales, torres2018partitioning}. In practice, however, these observations are often highly sparse and irregular in space due to orbital sampling constraints and environmental factors such as cloud cover, with effective coverage that can be extremely limited at any given time. Despite this sparsity, satellite measurements sample a dynamically coherent two-dimensional surface field that encodes information about the ocean interior through physical constraints such as geostrophic balance and stratification. This is starkly in contrast to the subsurface observations which consist of isolated vertical profiles and trajectories that do not form a spatially resolved three-dimensional field. Consequently, while surface observations can be interpreted as incomplete samples of an underlying structured field, the ocean interior remains intrinsically under-observed and must be inferred. This distinction motivates reconstruction approaches that leverage physically informed relationships between surface signals and subsurface structure, including early dynamical methods that project sea surface height into the ocean interior \cite{hurlburt1986dynamic, cooper1996altimetric}.

Traditional dynamical methods reconstruct subsurface ocean fields from surface observations, particularly satellite SSH, using a range of dynamical frameworks, including surface quasi-geostrophic (SQG-) based methods as well as broader model-based reconstructions \cite{isern2008reconstruction, fresnay2018reconstruction, liu2019reconstructing}. Hybrid dynamical-statistical approaches have also been developed, including those that combine SQG dynamics with multivariate empirical orthogonal function (EOF) reconstruction to retrieve interior density structure from surface data \cite{yan2020sqgmeofr}. These approaches often rely on simplified geostrophic dynamics, statistical projection modes, or reduced-order physical assumptions, which may limit their applicability in regions where ageostrophic processes are prominent. While these approaches have demonstrated encouraging results, their performance can degrade under high-resolution or extremely sparse observational regimes, and they are associated with high computational costs.

In recent years, machine learning and data assimilation methods (DA) have expanded the possibilities for the reconstruction of ocean interior states from surface observations. For example, Bolton and Zanna \cite{bolton2019deep} demonstrated, in a simplified quasi-geostrophic setting, that convolutional neural networks (CNNs) models can infer unresolved turbulent processes and even predict subsurface flow fields from degraded surface information alone. However, their framework focused on deterministic inference in simplified model dynamics rather than probabilistic reconstruction of full three-dimensional subsurface fields from extremely sparse real-world observations. Similarly, Meng et al. \cite{meng2022reconstructing} used CNNs to infer subsurface temperature and salinity anomalies using satellite-derived surface variables such as wind stress fields, sea surface temperature anomaly, sea surface salinity anomaly, and sea level anomaly. Their framework showed strong performance at high and super-resolution scales; however, its accuracy decreases as depth increases, as the statistical correlation between surface and deep ocean variability weakens. 

Generative AI approaches have been suggested as an alternative to the limitations of deterministic discriminative models. Recently, Martin et al. \cite{martin2025genda} developed a generative data assimilation framework for joint surface ocean state estimation from multimodal satellite observations, showing that diffusion-based priors can improve multivariable, multiscale surface reconstructions and infer unobserved surface currents. Similar generative approaches have also been proposed for super-resolution and inference from sparse and gappy observations in idealized quasi-geostrophic turbulence \cite{sureshbabu2025guided}. Although, in related work, Asefi et al. \cite{asefi2025glda} extended generative reconstruction to extremely sparse regional surface observations over the Gulf of Mexico, a setting with complex coastlines, bathymetry, and time-varying sampling, these studies remain focused on surface or idealized 2D settings and they do not address the reconstruction of full three-dimensional subsurface temperature, salinity, and velocity fields over several vertical levels.
More recently, Souza et al. \cite{souza2025surface} developed a score-based generative framework that can be used to reconstruct subsurface velocity and buoyancy statistics from only sea surface height (SSH). This framework provides a probabilistic estimate of the ocean state and associated uncertainty. However, the framework was not evaluated using sparse real-world observations or in dynamically complex coastal regions, such as semi-enclosed basins like the Gulf of Mexico, where boundary interactions, stratification variability, and multiscale processes introduce additional challenges \cite{kokoszka2023stratification, mckinney2021gulf, bracco2019mesoscale}.


In this paper, we propose a depth-aware conditional diffusion model to address the challenge of generating accurate and physically consistent three-dimensional subsurface fields from \textbf{extremely sparse} ($99.9\%$) SSH and ($73\%$) SST observations in the absence of any background dynamical model. While Souza et al. \cite{souza2025surface} proposed a related problem using a full-resolution SSH condition in the open ocean simulation settings, our contributions focus on real satellite observations and target the more complex regional dynamics of the Gulf of Mexico, characterized by intricate bathymetry, coastal boundaries, and highly sparse, time-varying measurements.

Notably, our depth-aware conditional DDPM generate subsurface fields at previously unseen depths within the same vertical range during inference. By providing continuous depth identifiers corresponding to new depths not included during training, we demonstrate that the model learns a continuous vertical representation rather than memorizing a discrete set of depth levels.

In order to measure the quality of the model, we use standard statistical evaluation metrics, including normalized root mean square error (NRMSE), correlation coefficient (CC), and the Structural Similarity Index (SSIM) between the estimated ocean states and reanalysis data from GLORYS~\cite{garric2018performance}, on which it is trained. SSIM compares the structural information, contrast, and luminance of two images to assess perceptual similarity. Unlike pointwise metrics such as MSE and CC, SSIM highlights structural dissimilarity between two images~\cite{nilsson2020understanding}.
Furthermore, to demonstrate the physical consistency of the estimated states for both small- and large-scale features, we inspect their Fourier spectrum as well as the predicted meridional heat transport.


\section{Results}
\label{sec:results}

In this section, we show the performance of our depth-aware conditional denoising diffusion probabilistic model (DDPM) in terms of statistical metrics such as NRMSE, CC, SSIM, as well as physics-based metrics such as the Fourier spectrum of the reconstructed variables, including temperature, salinity, zonal velocity, and meridional velocities, as well as the meridional heat-flux at $26^\circ$N. Notably, this reconstruction is performed in a zero-shot configuration, with no background dynamical model or ensemble-based data assimilation. We also compare this configuration of the DDPM model with a deterministic background conditioning using a Fourier neural operator (FNO+DDPM) and a UNet (UNet+DDPM).

\subsection{Reconstruction of subsurface ocean states from sparse surface observations}

Figure~\ref{reconstruct} presents the depth-aware DDPM reconstruction of subsurface temperature (T), salinity (S), and velocity fields (U, V) at three representative depths (55~m, 318~m, and 1062~m). The model is conditioned on highly sparse surface observations, e.g., SSH ( 99.9\% sparsity) and SST (73\% sparsity), as shown in panel (a), together with the log-normalized depth identifier used to specify each vertical level. Despite the severe sparsity of the conditioning inputs, the reconstructed fields recover the dominant spatial structures and their depth-dependent organization, with only minor deviations from the GLORYS reanalysis reference. Reconstructions at additional depths (25~m, 109~m, 155~m, 222~m, 453.9~m, and 643~m) are provided in Supplementary Figures S1 and S2, demonstrating consistent performance across the water column.

We further compare the depth-aware DDPM with hybrid models incorporating deterministic priors (FNO+DDPM and UNet+DDPM). Representative reconstructions at the same depths are shown in Supplementary Figures~S7--S9. While all models capture the large-scale structure of the subsurface fields, the hybrid variants introduce increased small-scale variability, some of which manifests as spurious features relative to GLORYS. Additional comparisons with the standalone depth-aware FNO and UNet models (Supplementary Figures~S10 and S11) show that deterministic architectures alone fail to recover the full structure of the subsurface fields with comparable fidelity. Taken together, these results highlight the advantage of the depth-aware probabilistic DDPM framework in reconstructing coherent three-dimensional ocean states from extremely sparse surface observations.

\subsection{Quantitative and physical evaluation of reconstruction performance}
\begin{figure}
  \centering
  \includegraphics[width=1\textwidth]{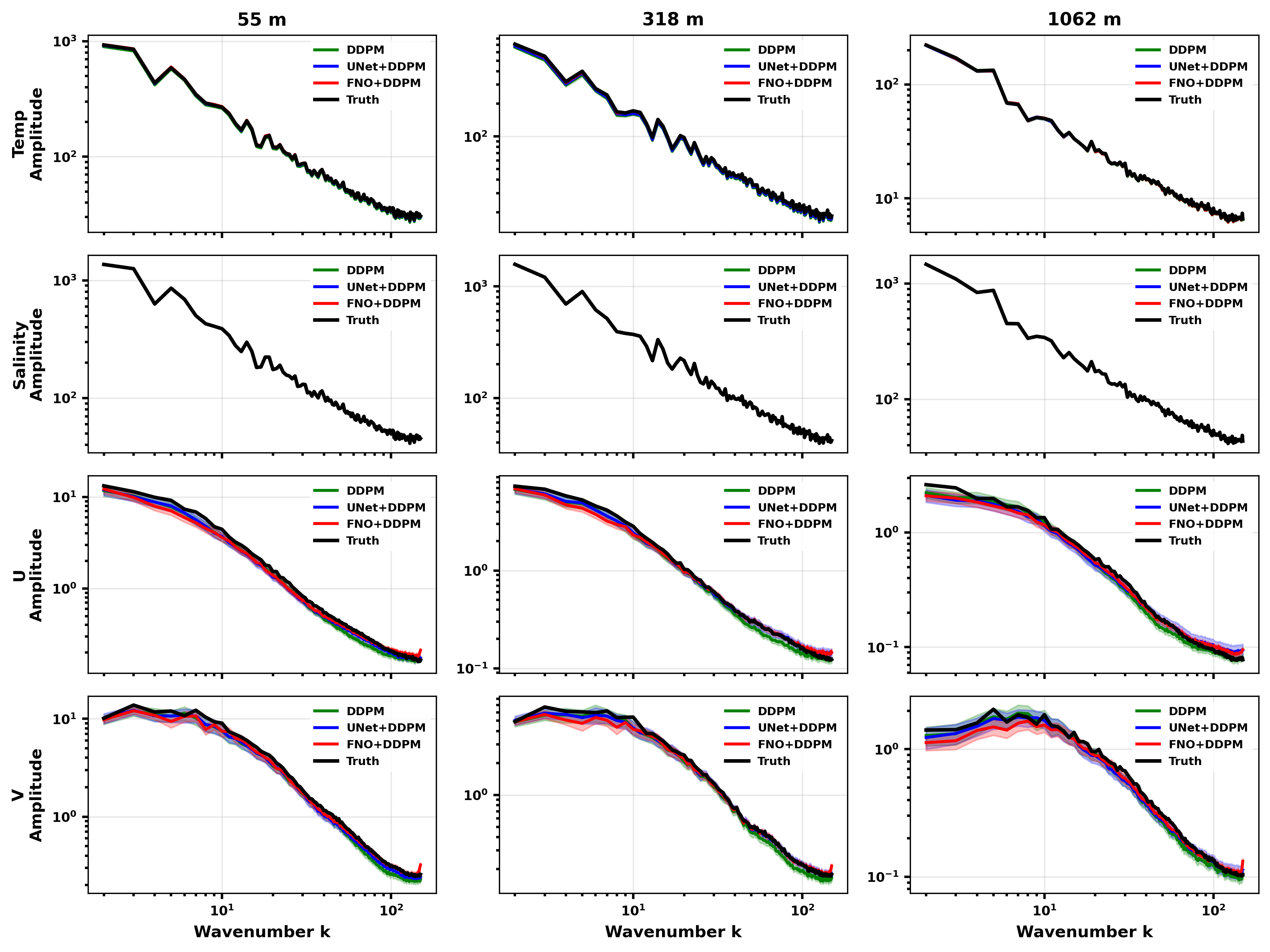}
     \caption{\textbf {Averaged Fourier power spectra for subsurface ocean variables at three depths.} 
     The power spectra of T, S, U, and V is displayed for 100 test samples at three different depths: 55 m, 318 m, and 1062 m.  Power spectra from the DDPM reconstructions are compared with the corresponding GLORYS reanalysis reference across spatial wavenumbers.}
  \label{spectrum}
\end{figure}

Figure~\ref{spectrum} shows the mean Fourier power spectra of T, S, U, and V computed across 100 test samples at three depths: 55~m, 318~m, and 1062~m. The depth-aware DDPM captures precise T and S spectra across all wavenumbers, showing successful reconstruction of both large-scale and small-scale features, even at the deepest level (1062 m). In contrast, velocity components are more sensitive to small-scale variability, resulting in slight variations at higher wavenumbers. Overall, these findings show that the depth-aware DDPM can effectively recover multi-scale dynamical structures over a variety of subsurface variables and depths. In addition, averaged Fourier power spectra at other depths are provided in the Supplementary Information, including estimates at 25~m, 109~m, and 155~m in Supplementary Figure S3 and at 222~m, 453.9~m, and 643~m in Supplementary Figure S4. These results show the same overall pattern: the velocity components are more sensitive to small-scale variability than T and S. Even at deeper levels, such as 643~m, the model still captures most of the large-scale and small-scale dynamics.

\begin{figure}
  \centering
  \includegraphics[width=1\textwidth]{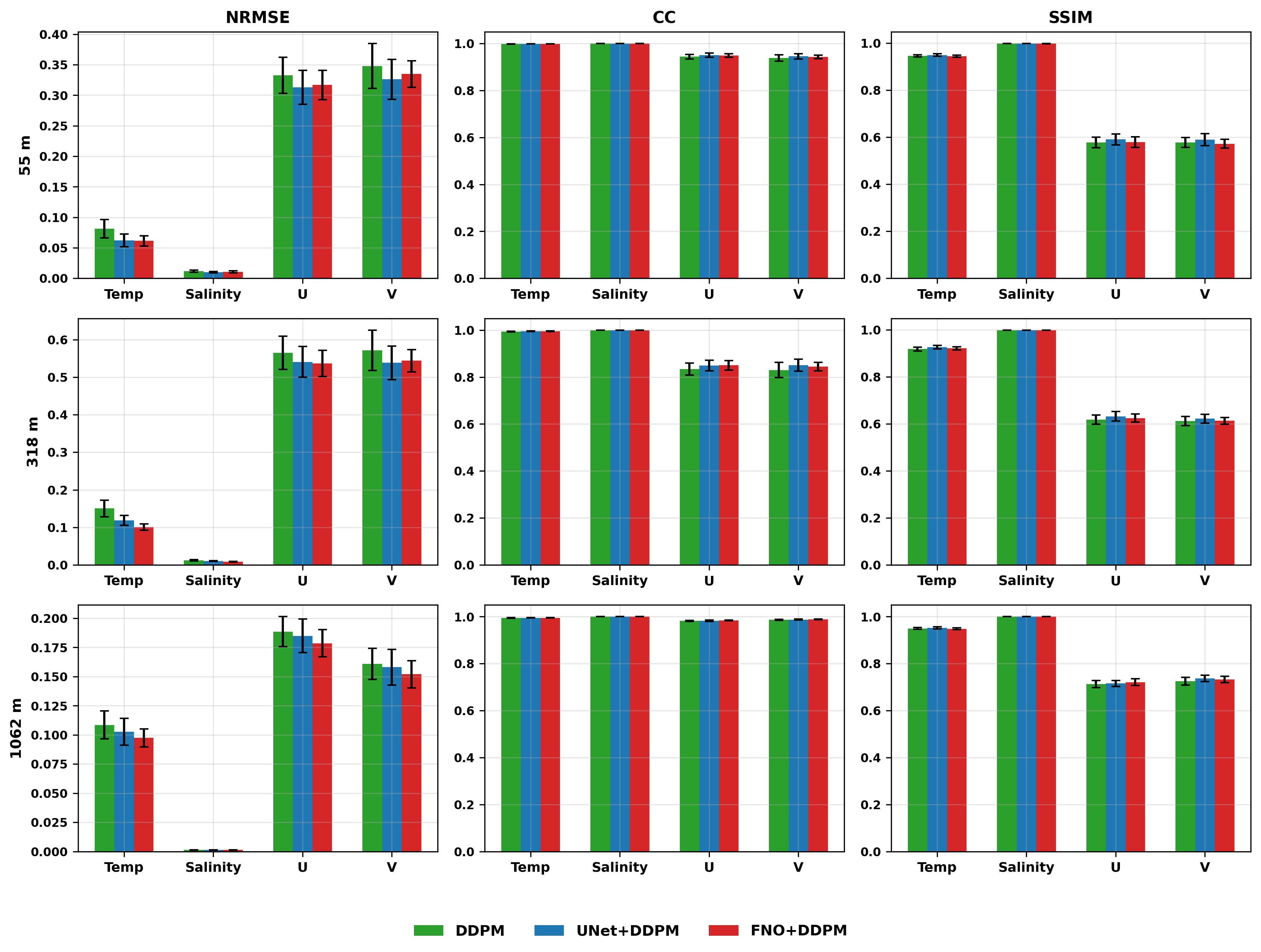}
  \caption{\textbf{Quantitative evaluation of reconstructed subsurface variables across depths and models.}
  Reconstruction performance for T, S, U, and V across 100 test samples. Columns correspond to various evaluation metrics, including normalized RMSE (NRMSE), correlation coefficient (CC), and SSIM. Results are compared across three models: DDPM, UNet+DDPM, and FNO+DDPM. Rows show results at three depths: 55\,m, 118\,m, and 1062\,m.}
  \label{std_metrics}
\end{figure}

Figure~\ref{std_metrics} demonstrates the reconstruction performance for T, S, U, and V across 100 test samples. Columns correspond to three evaluation metrics—normalized RMSE (NRMSE), CC, and SSIM—evaluated at three depths: 55~m, 318~m, and 1062~m. NRMSE is used to enable a fair comparison across variables with different physical units and magnitude ranges, including T, S, and velocity fields. The same quantitative metrics for additional depths are provided in the Supplementary Information, including results at 25~m, 109~m, and 155~m in Figure~S5, and at 222~m, 453.9~m, and 643~m in Figure~S6.

The results are presented for three models: DDPM, UNet+DDPM, and FNO+DDPM. Across all depths, the models generate S and T more accurately than velocity components, as reflected by lower NRMSE and higher CC and SSIM values. This behavior is expected and consistent with the Fourier spectra and snapshots , because velocity fields comprise more small-scale variability, which is more difficult to recover than the comparatively smoother T and S fields.

To evaluate whether incorporating a learned background prior improves performance of our depth-aware DDPM model, both UNet \cite{ronneberger2015unet} and Fourier Neural Operator (FNO) ~\cite{li2021fourier} baseline models were trained independently using the same depth-aware architecture and depth identifiers, and their outputs were subsequently used to condition the DDPM (UNet+DDPM and FNO+DDPM). It is important to note that conventional UNet and FNO architectures, without the proposed depth-aware modification, are inherently unable to recover three-dimensional ocean dynamics from extremely sparse observations. In Asefi et al. \cite{asefi2025glda}, we showed that, even for surface-field reconstruction, standard UNet and FNO models fail to accurately recover fine-scale structures and high-wavenumber dynamics when trained on $99\%$-$99.9\%$ sparse inputs.
While modest improvements are observed, particularly for FNO+DDPM, which achieves the lowest overall errors, the performance gains relative to the depth-aware DDPM are limited. This implies that the primary factor enabling accurate subsurface reconstruction for different depth layers is the incorporation of explicit depth identifiers, which allow the model to generalize effectively across depths, instead of reliance on an external background model.

\begin{figure}
  \centering
  \includegraphics[width=1\textwidth]{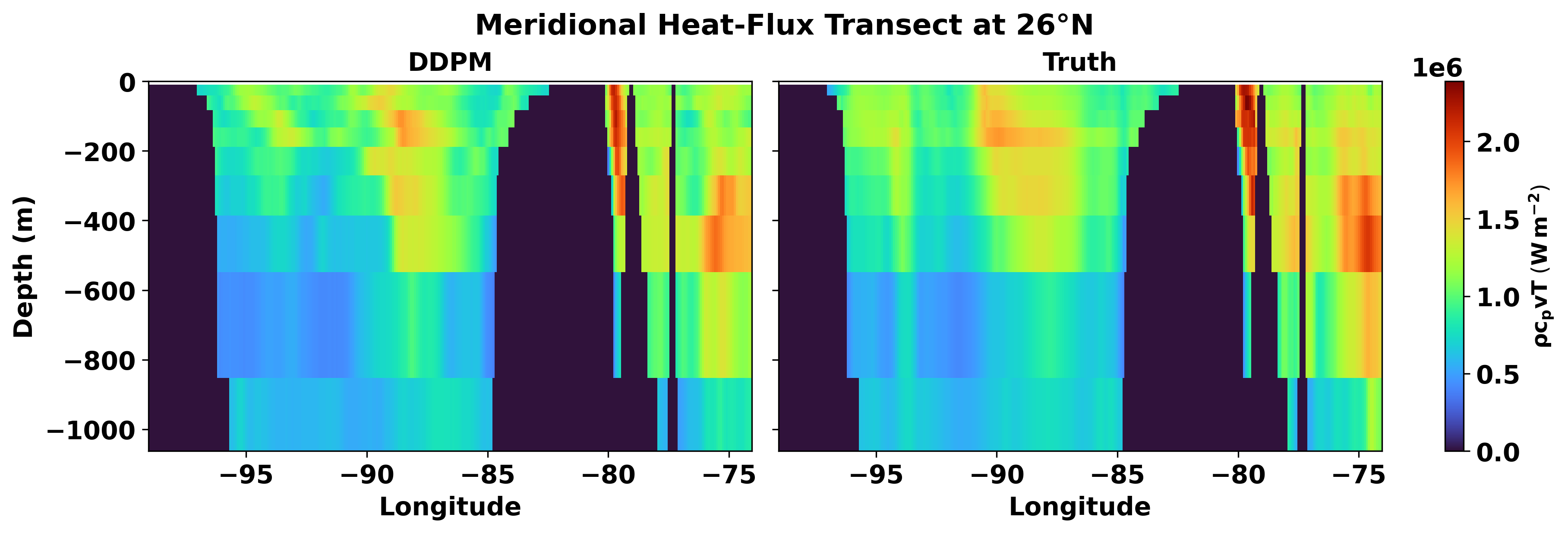}
  \caption{\textbf{Meridional heat-flux longitude--depth transects at $26^\circ$N on 13 February 2023.}
  Longitude--depth transects of meridional heat flux for the GLORYS reanalysis dataset (Truth) and the DDPM reconstruction are shown over the longitude range $[-99^\circ, -74^\circ]$.}
  \label{heatflux_metrics}
\end{figure}

Figure~\ref{heatflux_metrics} shows longitude--depth transects of the meridional ocean heat flux at $26^\circ$N, a latitude intersecting major circulation features in the Gulf of Mexico, to evaluate the physical consistency of the reconstructed subsurface fields. We examine the meridional ocean heat flux defined as
\begin{equation}
q_v = \rho c_p V T,
\end{equation}
where $\rho$ is the density, and $c_p$ is the heat capacity. This diagnostic metric characterizes the vertical and zonal distribution of oceanic heat transport. The DDPM reconstruction captures the magnitude, vertical structure, and sign variations of the meridional heat flux relative to the GLORYS reanalysis, demonstrating that the model preserves physically meaningful heat transport pathways beyond pointwise reconstruction accuracy.

\subsection{Zero-shot generalization to unseen depths}
\begin{figure}
  \centering
  \includegraphics[width=1\textwidth]{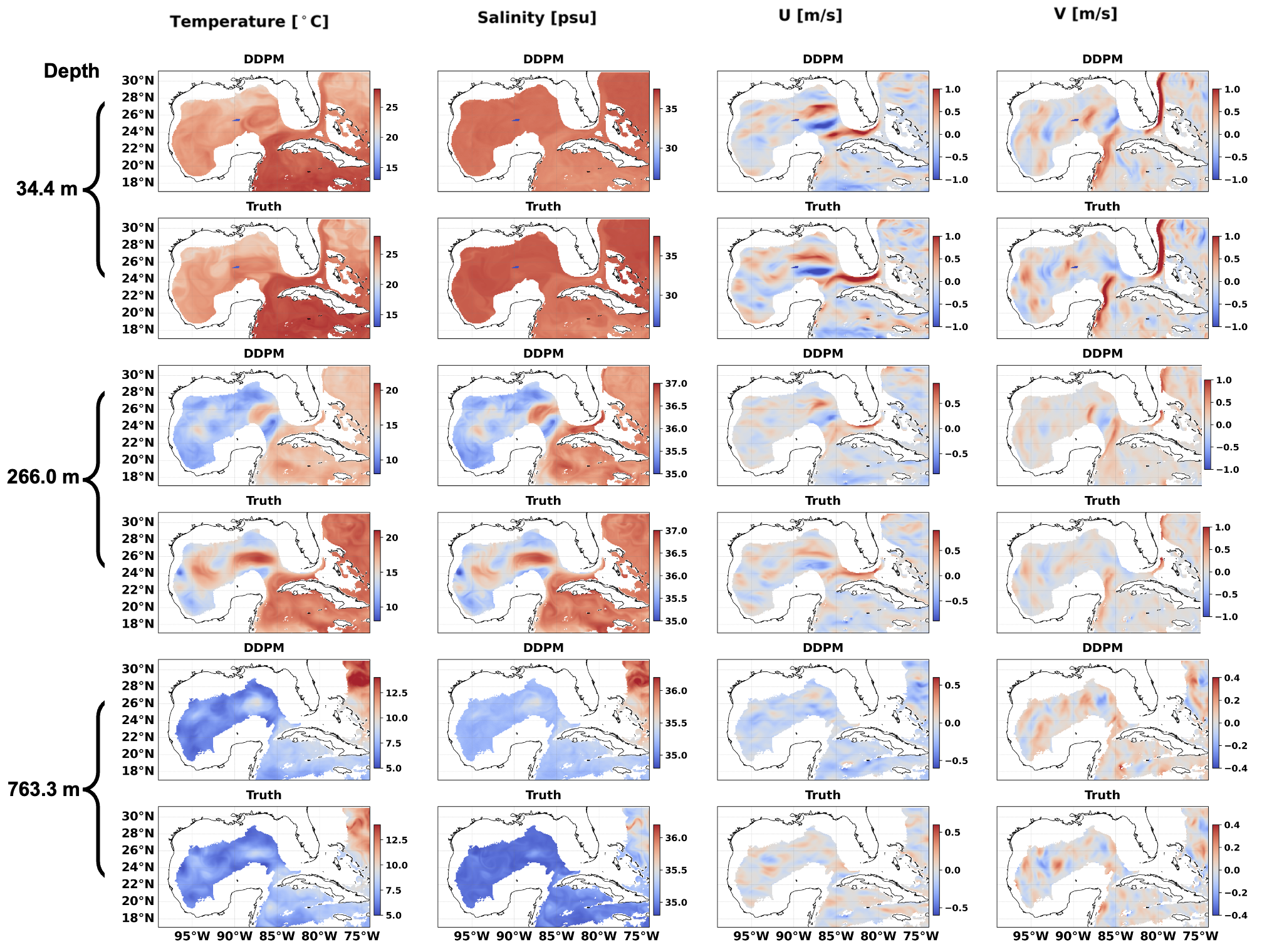}
  \caption{\textbf{Depth-aware DDPM reconstructions at three previously unseen depths within the training vertical range.} Reconstructions are shown for previously unseen depths of 34.43\,m, 266\,m, and 763.3\,m, which were not included during training.}
  \label{interpolate_Ddpm}
\end{figure}

To assess whether the model learns meaningful vertical structure when trained on nine discrete depth levels, we evaluate its ability to generalize to previously unseen depths by providing new depth identifiers during sampling within the DDPM architecture. In particular, three depths which are not included during training are sampled. Figure~\ref{interpolate_Ddpm} presents snapshot reconstructions of T, S, U, and V at these unseen depths such as 34.43\,m, 266\,m, and 763.3\,m,. The results show that large-scale spatial structures are well captured, while slightly larger deviations are observed compared to samples at trained depths shown in Figure~\ref{reconstruct}, which is expected. In general, the model generates subsurface fields with full resolution and exhibits that it has acquired a continuous depth representation instead of memorizing a discrete set of depth levels.

\begin{figure}
  \centering
  \includegraphics[width=1\textwidth]{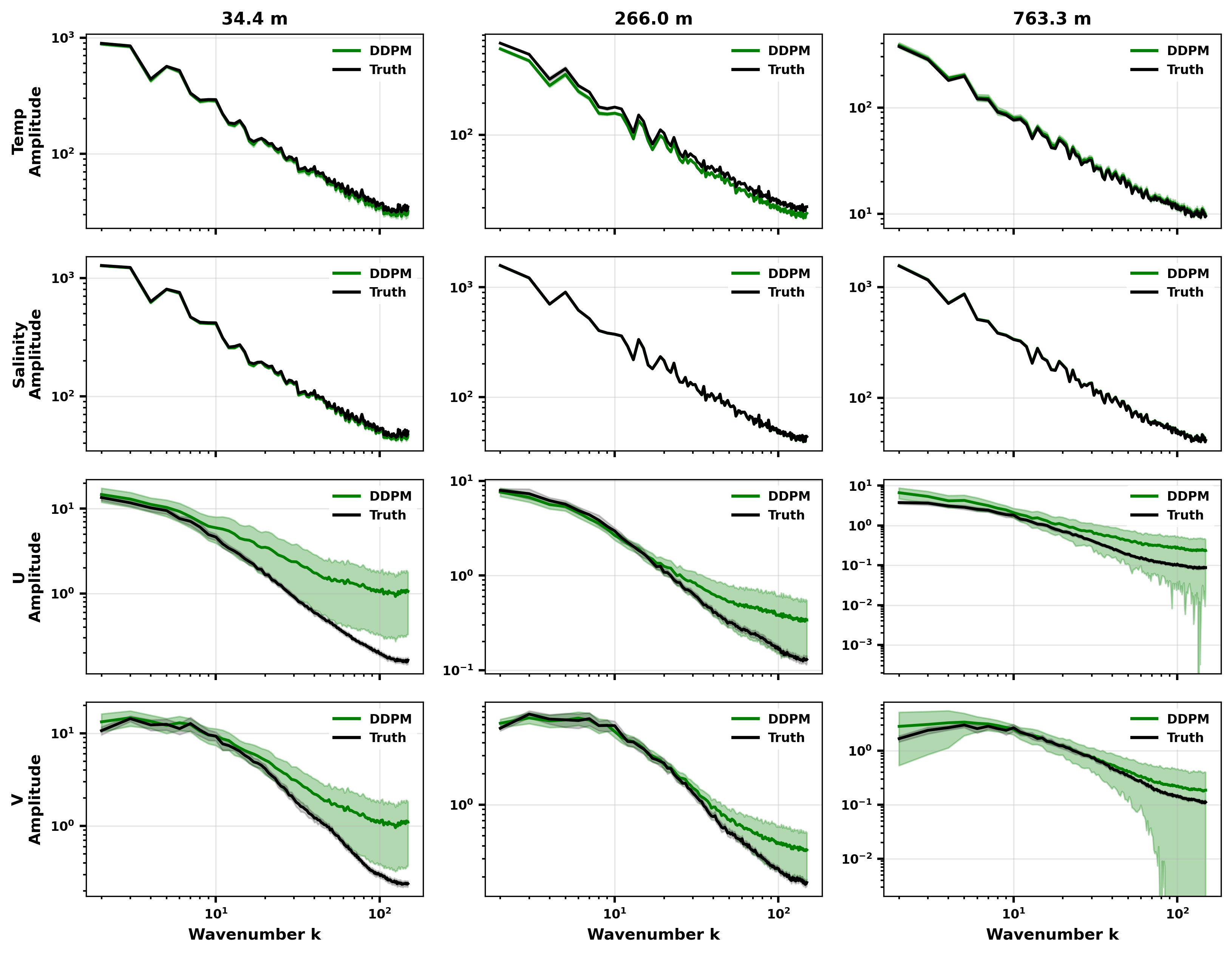}
  \caption{\textbf{Averaged Fourier power spectra for subsurface variables at previously unseen depths.} Power spectra for temperature, salinity, zonal velocity (U), and meridional velocity (V) are shown for 100 test samples at depths of 34.43\,m, 266\,m, and 763.3\,m that were not used during training.}
  \label{interpl_Spectm}
\end{figure}

Figure~\ref{interpl_Spectm} presents the averaged Fourier power spectra of T, S, U, and V at three previously unseen depths: 34.43~m, 266~m, and 763.3~m. T and S spectra are dominated by large-scale variability, and the DDPM reconstruction closely matches the ground truth GLORYS spectra across all wavenumbers, which indicates a physically consistent reconstruction even at unseen depth levels within the training vertical range. 

In contrast, the velocity components show larger deviations from the reference spectra and show stronger small-scale variability. At shallower unseen depths (34.43~m and 266~m), the DDPM accurately captures the large-scale spectral structure, while divergences increase at higher wavenumbers, which correspond to smaller-scale dynamics. At the deepest unseen depth (763.3~m), deviations become more noticeable, particularly at higher wavenumbers, reflecting the increased difficulty of reconstructing fine-scale velocity fluctuations at depth. Overall, these results suggest that the model generalizes well across unseen depths at shallower levels, preserving the dominant multi-scale structure while exhibiting expected limitations in small-scale velocity reconstruction while collapsing at deeper levels of the ocean.

\begin{figure}
  \centering
  \includegraphics[width=1\textwidth]{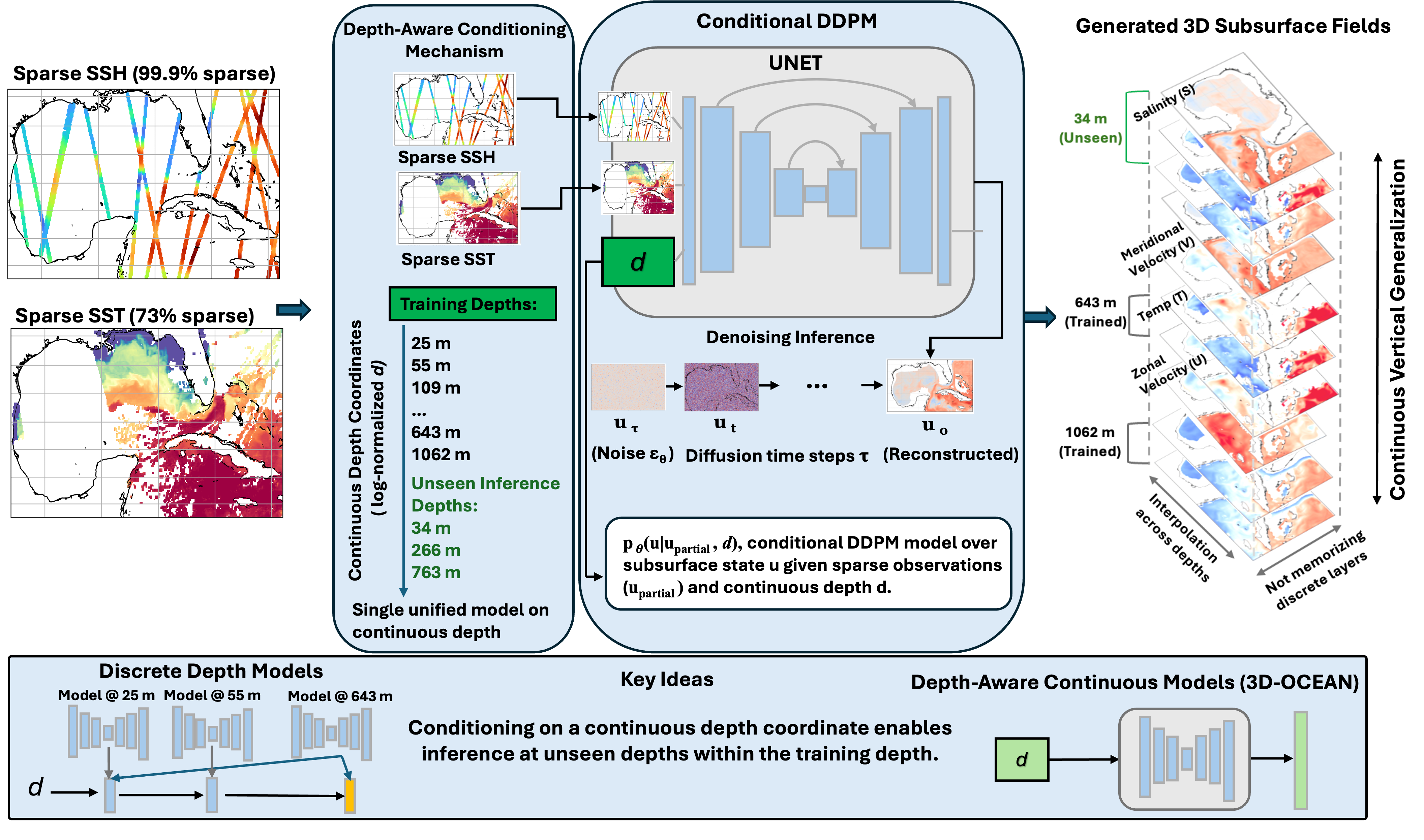}
  \caption{\textbf{Schematic of the depth-aware conditional DDPM framework for three-dimensional ocean reconstruction.}
Sparse observations of SSH and SST are used with a log-normalized continuous depth coordinate and used as conditioning inputs to a conditional denoising diffusion probabilistic model (DDPM). A single unified model is trained across nine discrete depth levels and learns a continuous vertical representation of the ocean interior. During inference, the model reconstructs T, S, U, and V fields not only at trained depths but also at previously unseen depths within the training range, enabling vertical interpolation without memorizing discrete layers.}
  \label{reconstruct}
\end{figure}

\begin{figure}
  \centering
  \includegraphics[width=1\textwidth]{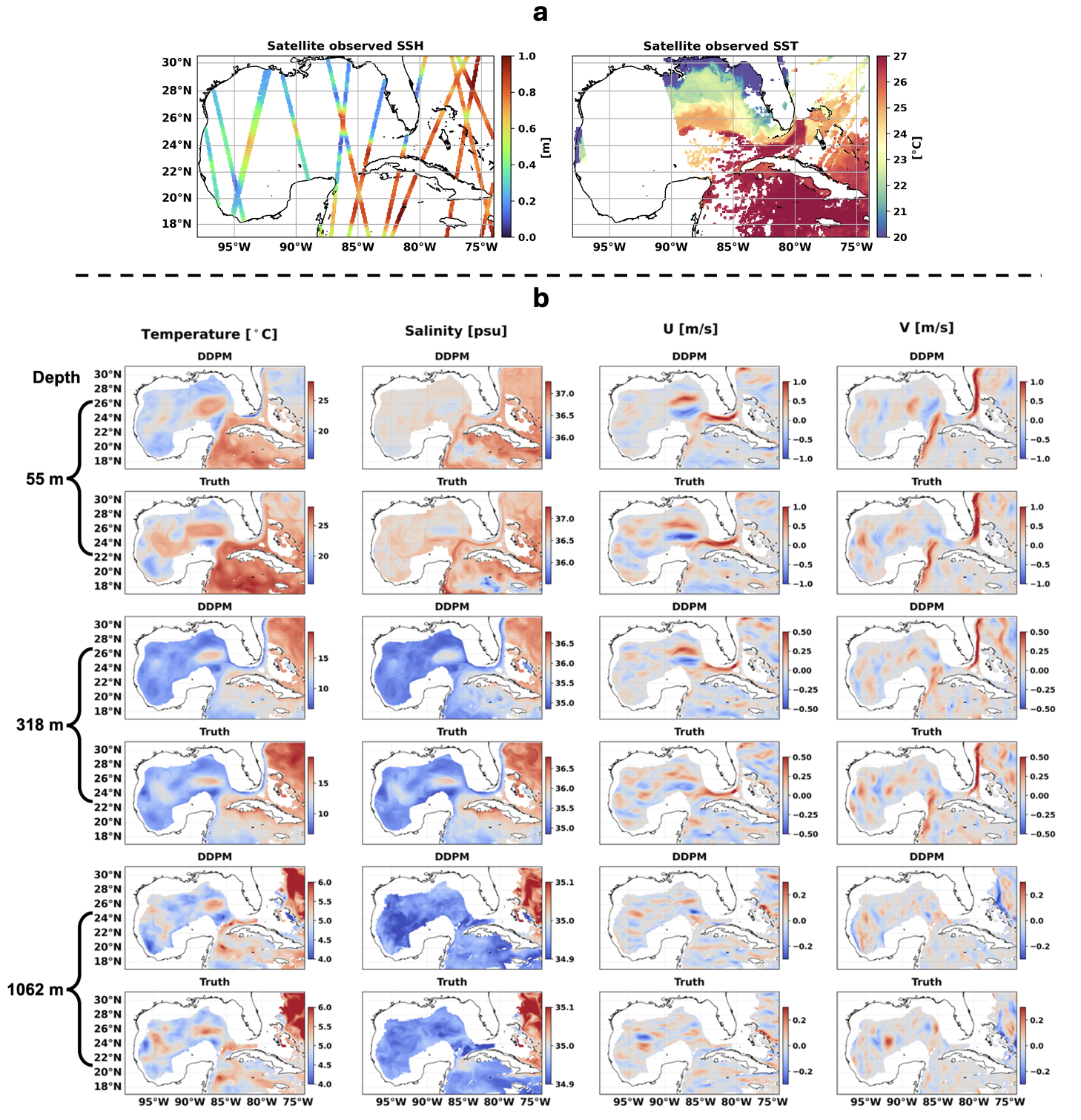}
  \caption{\textbf{Sparse surface observations and depth-aware DDPM reconstruction of subsurface ocean variables at three depths.} Panel a shows sparse satellite SSH and SST observations on February 13, 2023, which are used to reconstruct 3D ocean states in panel b.
     Panel b shows subsurface ocean variables, including T, S, U, and V is shown over the Gulf of Mexico at three different depths: 55~m, 318~m, and 1062~m. The DDPM reconstructions (top row) are compared with the corresponding ground-truth fields from the GLORYS reanalysis (bottom row) for each depth. Only three depths are displayed here for clarity, and reconstructions at the other depths are included in the Supplementary Material.  
White areas indicate land/bathymetry. As depth increases, the valid ocean domain decreases due to bathymetric constraints, resulting in a larger land mask at deeper levels.}
  \label{reconstruct}
\end{figure}

\section{Discussion}

In this work, we address a central limitation in ocean observation. The three-dimensional ocean state is not directly observable and must be inferred from sparse and heterogeneous measurements. In situ platforms provide high-fidelity vertical profiles, but their spatial and temporal sampling is limited. Satellite observations provide broader coverage, yet are restricted to surface variables. This leads to an underdetermined problem in which the ocean interior must be reconstructed from incomplete projections of the underlying dynamics. Our results show that a depth-aware DDPM provides a viable and scalable framework for reconstruction under conditions of extreme observational sparsity.

A key advance of this approach lies in its treatment of the vertical dimension. Existing methods typically rely on fixed depth discretizations or depth-specific mappings, which assume that vertical structure can be represented as independent layers. In contrast, the use of continuous, log-normalized depth identifiers enables the DDPM to learn a unified representation of the ocean interior across depths. This allows the model to generalize beyond the discrete levels used during training and generate consistent reconstructions at previously unseen depths. The ability to interpolate in depth space, rather than memorize depth-specific relationships, represents a conceptual shift in modeling three-dimensional ocean structure from surface observations.

The results further show that probabilistic generative modeling offers clear advantages over deterministic approaches in this setting. Across all evaluated depths, the DDPM reconstructs large-scale circulation patterns while preserving multiscale variability. This is supported by pointwise metrics such as NRMSE, CC, structural metrics like SSIM, as well as Fourier spectral agreement, which indicate that the model captures the distribution of energy across spatial scales. Deterministic baselines, in contrast, either oversmooth the fields or introduce spurious small-scale features, particularly in velocity components. These differences highlight the importance of representing uncertainty and multimodality when reconstructing underconstrained geophysical systems.

Beyond statistical measures, the preservation of physically meaningful diagnostics provides stronger evidence of model fidelity. The agreement in meridional heat-flux longitude–depth transects at 26°N shows that the reconstructed fields remain dynamically coherent and capture the structure and sign of heat transport pathways. These diagnostics depend on coupled interactions between temperature and velocity fields and are therefore sensitive to both amplitude and phase errors. The ability of the DDPM to recover these structures indicates that it learns representations consistent with underlying physical processes, even without explicit dynamical constraints.

At the same time, the results reveal important limitations. While temperature and salinity are reconstructed with high fidelity across depths, velocity fields exhibit larger deviations, particularly at smaller spatial scales and at previously unseen depths. This reflects both the increased complexity of velocity dynamics and the weaker coupling between surface observations and interior flow with depth. Addressing these limitations will require incorporating additional physical constraints into the generative framework, such as conservation laws, dynamical priors, or spectral regularization during training.

More broadly, this work suggests a new paradigm for ocean state estimation in data-limited regimes. Instead of relying solely on computationally expensive data assimilation systems or deterministic inversion methods, generative models such as DDPM can learn physically informed priors directly from data and perform reconstruction in a probabilistic and scalable manner. This creates opportunities for hybrid approaches in which generative models provide initial conditions, uncertainty estimates, or surrogate representations within forecasting systems.

Future work will focus on integrating DDPM-based reconstructions into dynamical forecasting frameworks such as ROMS and data-driven models such as OceanNet~\cite{chattopadhyay2024oceannet}. Evaluating forecast skill, error growth, and energy transfer across scales from DDPM-initialized states will provide a more stringent assessment of physical realism and predictive capability. In parallel, incorporating physics-aware constraints into the diffusion objective will be essential for improving dynamical consistency, particularly under extreme observational sparsity and in regimes where surface–subsurface coupling is weak.

\section{Dataset}
\subsection{GLORYS subsurface reanalysis data}
We use subsurface ocean state variables from the GLORYS reanalysis dataset~\cite{garric2018performance}, including temperature (T), salinity (S), zonal velocity (U), and meridional velocity (V), over the Gulf of Mexico at $\frac{1}{12}^{\circ}$ horizontal resolution.
 This reanalysis data is obtained by integrating observations with the NEMO ocean model~\cite{storkey2010forecasting}. We used a total of $11000$ temporal samples of T, S, U, and V from January $1993$ to February of $2023$ for training. We report the performance of our proposed model on $100$ independent temporal samples from outside the training set. 
\subsection{Satellite surface observations}
As conditioning inputs to our DDPM model, we use real satellite-derived surface observations of SSH and SST. These observations are inherently sparse and time-varying due to the orbital sampling pattern of satellites, with approximately $99.9\%$ spatial sparsity for SSH and $73\%$ for SST. The observations come from multiple altimetry missions since January 1, 1993, and exhibit strong spatial sparsity and highly non-uniform spatio-temporal sampling due to nadir-only measurement geometry. The SSH data are sourced from CMEMS and processed through DUACS, including inter-mission homogenization and corrections for atmospheric effects, tidal signals, and long-wavelength errors. The daily data are converted into super-observations by spatial averaging within each GLORYS grid cell, and a total of $11000$ samples from $1993$ to $2023$ are used for training.

\section{Methods}
We present a framework for reconstructing high-resolution 3D ocean states comprising T, S, U, and V, from sparse surface observations of sea SSH and SST. Our proposed model is a depth-aware conditional Denoising Diffusion Probabilistic Model (DDPM) which builds on the original DDPM framework introduced by Ho et al.~\cite{ho2020ddpm} while incorporating depth identifier and sparse surface observations as conditioning inputs for 3D ocean-state reconstruction. To assess its performance, we also implement two deterministic baselines adapted for depth-conditioned reconstruction: a depth-aware UNet and a depth-aware Fourier Neural Operator (FNO). In addition, we investigate two hybrid architectures, UNET+DDPM and FNO+DDPM, which use deterministic model predictions as priors to guide the diffusion-based reconstruction process.

\subsection{ Depth-aware conditional DDPM}

We introduce a depth-aware conditional DDPM designed to reconstruct high-resolution 3D ocean dynamics--- T, S, U, and V---from sparse surface observations of SSH and SST. The model is trained on nine discrete vertical levels: [25.21~m, 55.76~m, 109.73~m, 155.85~m, 222.48~m, 318.13~m, 453.94~m, 643.57~m, and 1062.44~m]. To enable a unified vertical representation, we encode each depth as a continuous normalized scalar conditioning variable. This formulation allows the model to learn a global vertical manifold rather than independent depth-specific mappings, thereby significantly reducing memory overhead compared to multi-model ensembles.

A critical component of our framework is the log-normalization of the depth identifier $d$. 
Although our subsurface and surface observations are normalized using min–max normalization, resulting in data within the range [0,1], the depth identifiers are normalized using a log-normalization equation to avoid any ``boundary bias'' in learning the continuous vertical range of our three-dimensional ocean model. When we applied min–max normalization, the shallowest layer (25 m) was mapped to zero, and the model failed to properly learn this level. In contrast, logarithmic normalization reduces this boundary bias by distributing depth values more uniformly in the normalized space, allowing the depth-aware conditional model to generalize consistently across all nine subsurface layers.

The primary architectural innovation of this framework is the integration of continuous depth identifiers as conditioning inputs.
This framework enables a single diffusion model to simultaneously learn representations across all subsurface levels and reconstruct full-resolution fields for T, S, U, and V at each depth. Unlike prior methods that require training separate models per depth or rely on fixed vertical discretizations, our framework supports continuous depth conditioning within a single generative model, which is memory-efficient. 

Significantly, our proposed depth-aware conditional DDPM shows strong generalization across depth. Unlike traditional architectures constrained to fixed discretizations, the continuous nature of our conditioning variable allows for zero-shot reconstruction at arbitrary depths within the training range. We validate this "vertical interpolation" capability by sampling the model at unseen depths during inference.

\subsubsection{Problem formulation}
The objective is to approximate the conditional distribution of the subsurface ocean state
$\mathbf{u} \in \mathbb{R}^{C \times H \times W}$, where $C=\{T,S,U,V\}$,
given sparse surface observations $\mathbf{u}_{\text{partial}}$ and a
log-normalized depth scalar $d$: This is formulated as Eq.~\ref{conditional_dist} 

\begin{equation}
p_\theta\left( \mathbf{u} \mid \mathbf{u}_{\mathrm{partial}}, d \right),
\label{conditional_dist}
\end{equation}

where $p_\theta\left( \mathbf{u} \mid \mathbf{u}_{\mathrm{partial}}, d \right)$ denotes the conditional distribution learned by the model with parameters $\theta$, $\mathbf{u}$ denotes the full-resolution subsurface T, S, U, and V across a continuous vertical range (discretized into nine levels for training) from GLORYS. The term $\mathbf{u}_{\mathrm{partial}}$ represents the sparsely observed surface inputs and the corresponding surface land-ocean mask, and $d$ denotes a continuous, log-normalized depth identifier associated with the target subsurface level.

The depth identifiers are incorporated as conditioning variables, allowing the diffusion model to learn depth-dependent representations across multiple vertical levels. 

\subsubsection{Reverse diffusion process and inference}

Our depth-aware conditional DDPM reconstructs high-resolution subsurface ocean states by sampling from the learned conditional distribution using sparse surface observations and continuous depth identifiers. Specifically, the model generates full-resolution fields of temperature (T), salinity (S), zonal velocity (U), and meridional velocity (V) corresponding to each depth identifier provided at inference time.

The use of continuous, log-normalized depth identifiers enables the model to generalize beyond the discrete depth levels observed during training. As a result, the reverse diffusion process can generate physically consistent reconstructions at previously unseen depths within the training range. In our experiments, we evaluate this capability by sampling subsurface fields at unseen depths of 34.43\,m, 266\,m, and 763.3\,m, which are not included in the training set.

To start the reverse sampling procedure, we initialize the state at the final diffusion step using Eq.~\ref{eq:init_reverse}, which combines the conditioning input with Gaussian noise:

\begin{equation}
u_T = \sqrt{\bar{\alpha}_T} \, u_{\mathrm{cond}}
+ \sqrt{1 - \bar{\alpha}_T} \, \epsilon,
\quad \epsilon \sim \mathcal{N}(0, \mathbf{I}),
\label{eq:init_reverse}
\end{equation}
where $u_{\mathrm{cond}}$ denotes the conditioning input constructed from sparse surface observations and the associated depth identifier.

At each reverse diffusion step $\tau$, the sample is updated using the learned denoiser:
\begin{equation}
u_{\tau - 1}
= \frac{1}{\sqrt{\alpha_\tau}}
\left(
u_\tau
- \frac{1 - \alpha_\tau}{\sqrt{1 - \bar{\alpha}_\tau}}
\, \epsilon_\theta
\bigl(
u_\tau, \tau, u_{\mathrm{cond}}
\bigr)
\right)
+ \sigma_\tau z,
\label{eq:reverse_step}
\end{equation}
where $\epsilon_\theta$ denotes the noise predicted by the conditional denoising network parameterized by $\theta$, $\sigma_\tau = \sqrt{\beta_\tau}$, and $z \sim \mathcal{N}(0, \mathbf{I})$ for $\tau > 1$, with $z = 0$ for $\tau = 1$.

\subsubsection{Power-law noise schedule}
We use a power-law beta scheduler that generalizes the standard linear schedule to improve training stability under extreme sparsity, as well as reducing spectral bias during samples as has been theoretically demonstrated in Sambamurthy et al.~\cite{sambamurthy2025lazy}. The diffusion variance at timestep $\tau$ is shown in Eq.~\ref{eq:power_law},

\begin{equation}
\beta_\tau = \beta_{\text{start}} + \left(\beta_{\text{end}} - \beta_{\text{start}}\right)
\left(\frac{\tau}{T}\right)^{p}
\label{eq:power_law}
\end{equation}

where $p$ controls the rate of noise growth. In our training, we used $p=2$, 
$\beta_{\text{start}} = 10^{-4}$, $\beta_{\text{end}} = 0.015$, and $T = 1000$ diffusion time steps. 
The model is trained using the \texttt{AdamW} optimizer with a learning rate of $10^{-4}$ 
to minimize the $L^2$ loss between the predicted noise and the ground-truth Gaussian noise added during the forward diffusion process.
This learning rate enabled the model to converge within a realistic training time on the available hardware.

\subsubsection{Computational scaling}

An important practical consideration in three-dimensional ocean reconstruction is the scaling of computational memory with respect to the number of vertical levels. In standard formulations, where each depth level is treated as a separate input channel or where models are trained independently across discrete depths, the input dimensionality grows linearly with the number of vertical levels $N_z$. As a result, both the memory footprint and computational cost of training scale as $\mathcal{O}(N_z)$, which becomes prohibitive for high-resolution three-dimensional fields.

In contrast, the depth-aware DDPM introduced in this work avoids this scaling by representing depth as a continuous conditioning variable rather than as an additional discrete dimension. Specifically, instead of stacking fields across $N_z$ levels, the model operates on two-dimensional spatial slices conditioned on a depth identifier $z$, allowing the same network to be reused across all depths. Consequently, the input dimensionality remains fixed, and the memory requirement does not increase with the number of vertical levels.

This distinction leads to a fundamentally different scaling behavior. While conventional approaches require memory that grows with $N_z$, the depth-aware formulation achieves constant memory complexity with respect to vertical resolution, that is, $\mathcal{O}(1)$ in $N_z$ for a fixed spatial grid. This property enables the extension of the framework to finer vertical discretizations and to continuous depth inference without incurring additional memory overhead.

From a practical standpoint, this allows training and inference on high-resolution three-dimensional domains that would otherwise be infeasible under standard architectures. More broadly, it highlights the advantage of treating depth as a continuous coordinate in generative modeling, rather than as a discrete stacking dimension, particularly in geophysical systems where vertical resolution is critical.

\subsection{Baseline deterministic models}

As noted by Souza \cite{souza2025surface} et al., deterministic models are fundamentally limited for subsurface reconstruction because they seek to infer a single subsurface state from surface observations, even though the mapping from the ocean surface to the interior is inherently ill-posed and non-unique. Moreover, standard deterministic architectures such as UNet and the FNO are primarily designed for 2D-to-2D mappings and are therefore not directly suited for reconstructing three-dimensional ocean dynamics across multiple depth levels. To address this limitation and provide a fair baseline comparison to our DDPM model, 
we adapt the framework to reconstruct 3D ocean dynamics by incorporating continuous, log-normalized depth identifiers as additional conditioning input.

\subsubsection{Depth-aware UNet baseline}
We implement a depth-aware UNet as a deterministic baseline. 
The input tensor $\mathbf{X} \in \mathbb{R}^{6 \times H \times W}$ comprises five surface channels---sparse Sea Surface Height (SSH), sparse Sea Surface Temperature (SST), their respective observation masks, and a land--ocean mask---augmented by a sixth channel containing the broadcasted log-normalized depth identifier $d$. The model is trained to minimize the MSE between the predicted and ground-truth fields for T, S, U, and V.

The encoder consists of two stages, each containing two $3 \times 3$ convolutional layers followed by ReLU activations and $2 \times 2$ max-pooling for spatial downsampling. The latent bottleneck consists of two $3 \times 3$ convolutions that transform 128 input channels into a 256-channel high-dimensional representation. The decoder is structurally symmetric to the encoder, utilizing $2 \times 2$ transposed convolutions with a stride of 2 to double the spatial resolution at each stage. After each upsampling operation, feature maps from the encoder are concatenated with the corresponding decoder layers through skip connections. These concatenated maps are then processed through convolutional blocks consisting of two $3 \times 3$ convolutions. The final layer employs a $1 \times 1$ convolution to project the feature maps into the four-channel output space, corresponding to the reconstructed fields of \{T, S, U, V\}. By training a single shared UNet architecture across all nine vertical levels, the network learns a unified mapping that treats depth as a continuous conditioning coordinate.

\subsubsection{Depth-aware FNO baseline}
We implement a depth-aware FNO as a secondary deterministic baseline, adapted for three-dimensional ocean-state reconstruction.
Neural operators $\mathcal{G}$ are effective for learning solution operators of partial differential equations (PDEs) and are designed to map an input function $\mathbf{X}$ to an output function $\mathbf{u}$ (i.e., $\mathcal{G}: \mathbf{X} \rightarrow \mathbf{u}$), a neural operator learns the functional relationship to predict $\mathbf{u} \in \mathcal{U}$ for any given $\mathbf{X} \in \mathcal{X}$, these models suffer from spectral bias, favoring low-wavenumber components and potentially challenging the recovery of fine-scale oceanographic structures~\cite{oommen2025integrating}.

FNOs operate in Fourier space by transforming the input, learning on a selected set of spectral modes, and then mapping the result back to physical space. Unlike standard convolutional networks, they capture long-range spatial dependencies and can generalize across different grid resolutions.

To enable 3D reconstruction, we augment the surface input $\mathbf{X} \in \mathbb{R}^{H \times W \times 5}$ with a sixth channel representing the broadcasted log-normalized depth scalar $d$. The architecture employs four Fourier layers with a width of 20, retaining 129 Fourier modes along each spatial dimension. The model is optimized using the Adam optimizer with a learning rate of $10^{-3}$ to minimize the Mean Squared Error (MSE):

\begin{equation}
\min_{\theta_{\mathrm{FNO}}} \mathbb{E} \left\| \mathbf{u}(x,y;d) - \mathcal{G}_{\theta_{\mathrm{FNO}}}(\mathbf{X}(x,y;d)) \right\|_2^2,
\end{equation}

where $\mathbf{u} \in \mathbb{R}^{H \times W \times 4}$ denotes the full-resolution target fields \{T, S, U, V\} at depth $d$.

\subsection{Hybrid conditioned variants: FNO+DDPM and UNet+DDPM}
Inspired by the findings in our previous work, Asefi et al. \cite{asefi2025glda}, we develop two hybrid-conditioned variants: UNet+DDPM and FNO+DDPM.
In these configurations, the diffusion model is conditioned not only on the raw sparse observations and log-normalized depth identifiers $(\mathbf{u}_{\text{partial}}, d)$ but also on the deterministic predictions of the pre-trained baseline models. In the UNet+DDPM variant, the conditioning inputs $\mathbf{u}_{\text{cond}}$ are concatenated with the deterministic output of the depth-aware UNet model. Similarly, in the FNO+DDPM variant, $\mathbf{u}_{\text{cond}}$ is augmented with the output of the depth-aware FNO. 

This hierarchical framework enables the DDPM to produce physically consistent, full-resolution reconstructions by incorporating the coarse outputs of the baseline models as learned physics-informed priors. As a result, each denoising step is guided by both the noisy input, depth identifier, and the large-scale spatial structure extracted from the sparse observations by the baseline models.

\section{Acknowledgement}
We thank Leonard Lupin-Jimenez for insightful discussions related to this project. AC and RH designed the research. NA conducted the research and developed the computational models. TW provided the GLORYS data, satellite observations, and heat transport diagnostics. NA and AC wrote the manuscript. All authors analyzed the results and edited the manuscript. Computational resources were provided by NCAR CISL UCSC0009, NSF ACCESS MTH260017, and National Energy Research Scientific Computing Center, a DOE Office of Science User Facility supported by the Office of Science of the U.S. Department of Energy under Contract No. DE-AC02-05CH11231 using NERSC award NERSC DDR-ERCAP0038781. NA and AC was supported by NSF grant no. 2425667. AC acknowledges the support from the Sloan Research Fellowship.

\section{Data and Code}
The computational models used in this study are publicly available at \url{https://github.com/TACS-UCSC/Ocean3D_Estimation}.

\section{References}
\makeatletter
\renewcommand{\refname}{\vspace{-\baselineskip}}
\makeatother
\bibliographystyle{unsrt}
\bibliography{references}

\end{document}